\documentclass{jpp}
\usepackage{subeqn}
\usepackage[dvips]{graphicx}
\ifprodtf \else
  \checkfont{eurm10}
  \iffontfound
    \IfFileExists{upmath.sty}
      {\typeout{^^JFound AMS Euler Roman fonts on the system,
                   using the 'upmath' package.^^J}%
       \usepackage{upmath}}
      {\typeout{^^JFound AMS Euler Roman fonts on the system, but you
                   dont seem to have the}%
       \typeout{'upmath' package installed. JFM.cls can take advantage
                 of these fonts,^^Jif you use 'upmath' package.^^J}%
      }
  \else
  \fi
\fi

\ifprodtf \else
  \checkfont{msam10}
  \iffontfound
    \IfFileExists{amssymb.sty}
      {\typeout{^^JFound AMS Symbol fonts on the system, using the
                'amssymb' package.^^J}%
       \usepackage{amssymb}%
         \let\leq=\leqslant

      }{}
  \else
  \fi
\fi

\ifprodtf \else
  \IfFileExists{amsbsy.sty}
    {\typeout{^^JFound the 'amsbsy' package on the system, using it.^^J}%
     \usepackage{amsbsy}}
    {\providecommand\boldsymbol[1]{\mbox{\boldmath $##1$}}}
\fi

\newcommand\bmath[1]{\ensuremath{\boldsymbol{#1}}}

\newsavebox{\astrutbox}
\sbox{\astrutbox}{\rule[-5pt]{0pt}{20pt}}

\newcommand{\sbk}[1]{\left[ #1 \right]}

\title[MD of Collision between a Graphite and a Hydrogen Atom]
 {Molecular Dynamics Simulation of Collisions between Hydrogen and Graphite }

\author[A.\,Ito and H.\,Nakamura]%
  {A\ls T\ls S\ls U\ls S\ls H\ls I\ns\ns I\ls T\ls O$^1$%
   \ns \and \ns H\ls I\ls R\ls O\ls A\ls K\ls I\ns\ns N\ls A\ls K\ls A\ls M\ls U\ls R\ls A$^2$}

\affiliation{$^1$Department of physics, Graduate School of Science, 
Nagoya University, Chikusa, Nagoya 464-8602, Japan\\[\affilskip]
$^2$National Institute for Fusion Science,
Oroshi-cho 322-6, Toki 509-5292, Japan}

\pubyear{2005}
\volume{*}
\part{*}
\pagerange{1--4}
\date{29 August 2005 and accepted 16 December 2005}
\begin{document}

\maketitle

\begin{abstract}
Hydrogen adsorption by graphite is examined by classical molecular dynamics simulation using a modified Brenner REBO potential. Such interactions are typical in chemical sputtering experiments, and knowledge of the fundamental behavior of hydrogen and graphene in collisional conditions is essential for modeling the sputtering mechanism. The hydrogen adsorption rate is found to be dependent on the incident hydrogen energy and not on graphene temperature. Rather than destroying the graphene, hydrogen incidence at energies of less than 100~eV can be classified into three regimes of adsorption, reflection and penetration through one or more graphene layers. Incidence at the lowest energies is shown to distort the graphene structure.
\end{abstract}

\section{Introduction}

In chemical sputtering experiments, 
 such as experiments using the diver tor of a nuclear fusion device, 
 plasma--carbon interactions have been reported to yield 
 hydrocarbon molecules of small number of atoms[\cite{Nakano}]. 
However, the mechanism of hydrocarbon generation has not been elucidated yet.
In the present study, this mechanism is examined in detail through simulation of the collision of hydrogen atoms with graphene as one of the fundamental processes in plasma--carbon interaction.
%During the collision process in the region where the incident hydrogen kinetic
%energy is less than 100 eV in our simulation, 
%it is expected that the hydrogen atom is absorbed by the graphenes, rather than that the hydrogen atoms destroy the graphene.

%In the present paper, we investigate collision mechanism of hydrogen 
%against the graphene  from a viewpoint of  `microscopic' atomic motion by simulation.

A classical molecular dynamics (CMD) simulation scheme
 is adopted in the present study to allow the dynamics of a many-particle system
 to be simulated over an adequate length of time given limited computer resources. 
Quantum mechanical simulations could not be performed for the same time
 of interaction using the resources available. 
A new model potential between hydrogen and graphene is introduced for these simulations
 in order to incorporate Brenner's REBO potential [\cite{rebo}] as the basis for the chemical reaction.
%and is improved to deal with  chemical reaction. 
%We, therefore, restrict incident particles to ground state hydrogen atoms.
The carbon wall is treated as graphene layers
 which are minimal structure of both a graphite and a carbon fiber.
However, for simplification, they do not inter-layer interaction.
The kinetic energy of incident hydrogen is set to less than 100~eV
 to facilitate comparison with the results of divertor experiments.

\section{Simulation model}

%Because a model of the inter-nuclear potential to deal with the all phenomena
% in the plasma-wall system does not exist, we restricted objects.
%Incident particles were limited to only neutral hydrogen atoms, 
%and we employed only one graphene as the carbon material of the wall 
% in order that the intermolecular force does not appear.
%The inter-nuclear potential has only to express covalent bonds and chemical reaction
% of only grand-state atoms.
%A potential function satisfying those requests was proposed as the REBO potential by Brenner[\cite{rebo}];
The interaction potential in CMD has been developed through contributions
 by many theorists [\cite{rebo}-\cite{Tersoff}].
%[\cite{rebo,Morse,Abell,Tersoff}].
%Based on their potentials, we propose the  modified Brenner's REBO potential.
The Brenner's original REBO potential [\cite{rebo}] has the following form:
\begin{eqnarray}
	U = \sum_{i,j(>i)} \sbk{V^R(r_{ij}) - \bar{b}_{ij}(\{\bmath{r}\}) V^A(r_{ij})},
\end{eqnarray}
where $r_{ij}$ is the distance between atoms $i$ and $j$,
 $V^A$ is an attractive term, $V^R$ is a repulsive term,
 and the function $\bar{b}_{ij}(\{\bmath{r}\})$ includes all the effects of molecular orbitals.
However, if chemical reactions occur,
 the REBO potential breaks energy conservation.
To deal with chemical reactions, new functions expressing conjugation effects are thus proposed:
\begin{eqnarray}
	N_{ij}^{conj} = 1 + \sum_{k(\neq i,j)}^{carbon} f^c(r_{ik})C_N(N_{ki}^t)
		 + \sum_{l(\neq j,i)}^{carbon} f^c(r_{jl})C_N(N_{lj}^t),
\end{eqnarray}
where $f^c$ is a cut-off function for the distance between atoms, and 
\begin{eqnarray}
	C_N(x) = \left\{ \begin{array}{ll}
				1 & {\rm if}\, x \leq 2 , \\
				\sbk{1+cos(\pi (x - 2))} / 2 & {\rm if}\, 2 < x \leq 3, \\
				0 & {\rm if}\, x > 3 ,  \\
\end{array} \right.  \\
	N_{ki}^t = \sum_{j(\neq k,j)} f^c(r_{kj}) - f^c(r_{ki}) = \sum_{j(\neq k,j,i)} f^c(r_{kj}) .
\end{eqnarray}
In contrast to Brenner's original formulation [\cite{rebo}], the second and third terms of $N_{ij}^{conj}$ are not squared. The tricubic spline functions $F$ and $T$ in Ref. \cite{rebo} are redefined, as the original functions have $N_{ij}^{conj}$ as a variable. Denoting a new function of $F$ by $F'$, we obtain $F'(i,j,4) = F(i,j,6)$, $F'(i,j,k \leq 5) = F(i,j,9)$ and $F'(i,j,k) =  F(i,j,k)$ for the other cases. The spline $T(i,j,k)$ in Ref. \cite{rebo} is redefined in a similar way.
%For this modification, the total energy is conserved
% as far as all functions forming $U$, cutoff function and spline function also,
% is faithfully differentiated to obtain the force functions.
The above modifications of $N_{ij}^{conj}$, $F'$ and $T'$ provide differentiability at the cut-off point, conserving the total energy in the chemical reaction. This modified potential is employed in the present CMD simulations.

Graphenes are set parallel to the $x-y$ plane in $z<0$ region.
Each graphene is comprised of 160 carbon atoms with a periodic boundary condition.
%A single graphene is set on the $x-y$ plane at $z=0$ as the `first' graphene,
% and is comprised of 160 carbon atoms with a periodic boundary condition.
The velocities of all atoms in the graphene are distributed according to
 a Maxwell's distribution at the graphene temperature. 
Incident hydrogen atoms arrive normal to the graphenes from the $z=4$~\AA .
%Here $x$ and $y$ are given as  random numbers.
%In the simulation, incidence of hydrogen is repeated using the initial conditions
The simulation continues until the incident atom is reflected, 
 the atom leaves the graphenes into the $z>0$ region,
 or until the atom becomes trapped by the graphenes.
%(We call  the graphene initialised  $n$-times to the `n-th' graphene.)

% the simulation renews the graphene with the kept velocity of the incident atom.
%after the incident atom penetrates through some graphenes and is reflected,
%it can also penetrate through first graphene again.
%That case means going out of the surface, and the simulation is finished.

A total of 200 simulations was performed for each graphene temperature and incident energy, with the initial position of the hydrogen atom on the $z=4$~\AA \ plane varied randomly to obtain statistical information.
%The above simulation was repeated 200 times per temperature and incident energy,
% while changing the initial position of incident atom.
The present CMD simulation satisfies the constant-volume-energy (NVE) condition.

%The simulation employs following scheme; 
% the graphene which is composed by 160 carbon atoms
% is set parallel to the x-y plane, 
%the boundary condition of x and y is periodic and
%the incident hydrogen atoms come from a positive direction of z axis.
%The simulation finishes when the incident atom goes out of the surface
% which means most.
%The simulation changing the initial position of atoms
% is done 200 times by the temperature of the graphene and incident energy.

\begin{figure}
\centering

\begin{tabular}{cc}
	\resizebox{60mm}{!}{\includegraphics{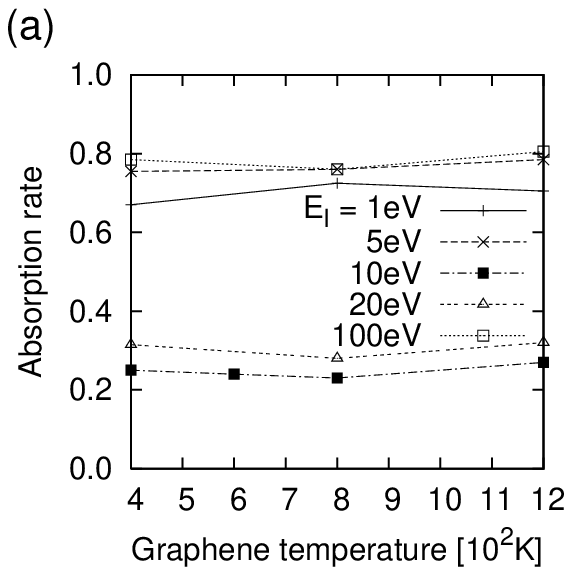}} &
	\resizebox{60mm}{!}{\includegraphics{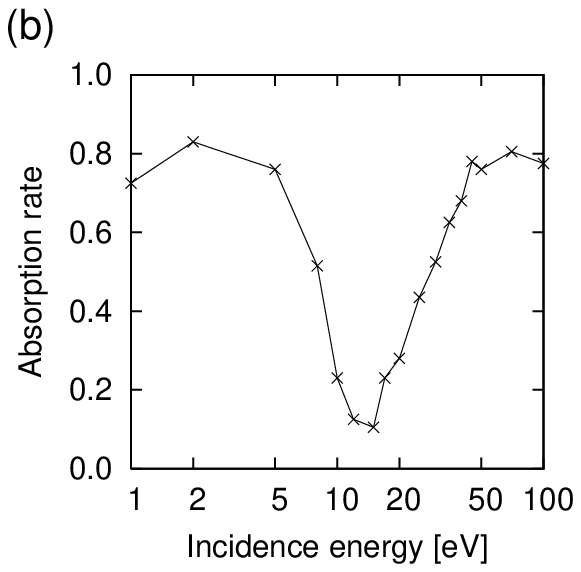}} \\
\end{tabular}
\caption{(a) Hydrogen adsorption rate vs. graphene temperature, and (b) incidence energy at 800~K }

%\includegraphics{4-Ito_fig2.eps}
%\caption{a hydrogen absorption rate as a function of a incident energy
% when the temperature of the graphene is 800 K.}

\label{fig:fig1}
\end{figure}

\begin{figure}
\centering
\begin{tabular}{cc}
		\resizebox{60mm}{!}{\includegraphics{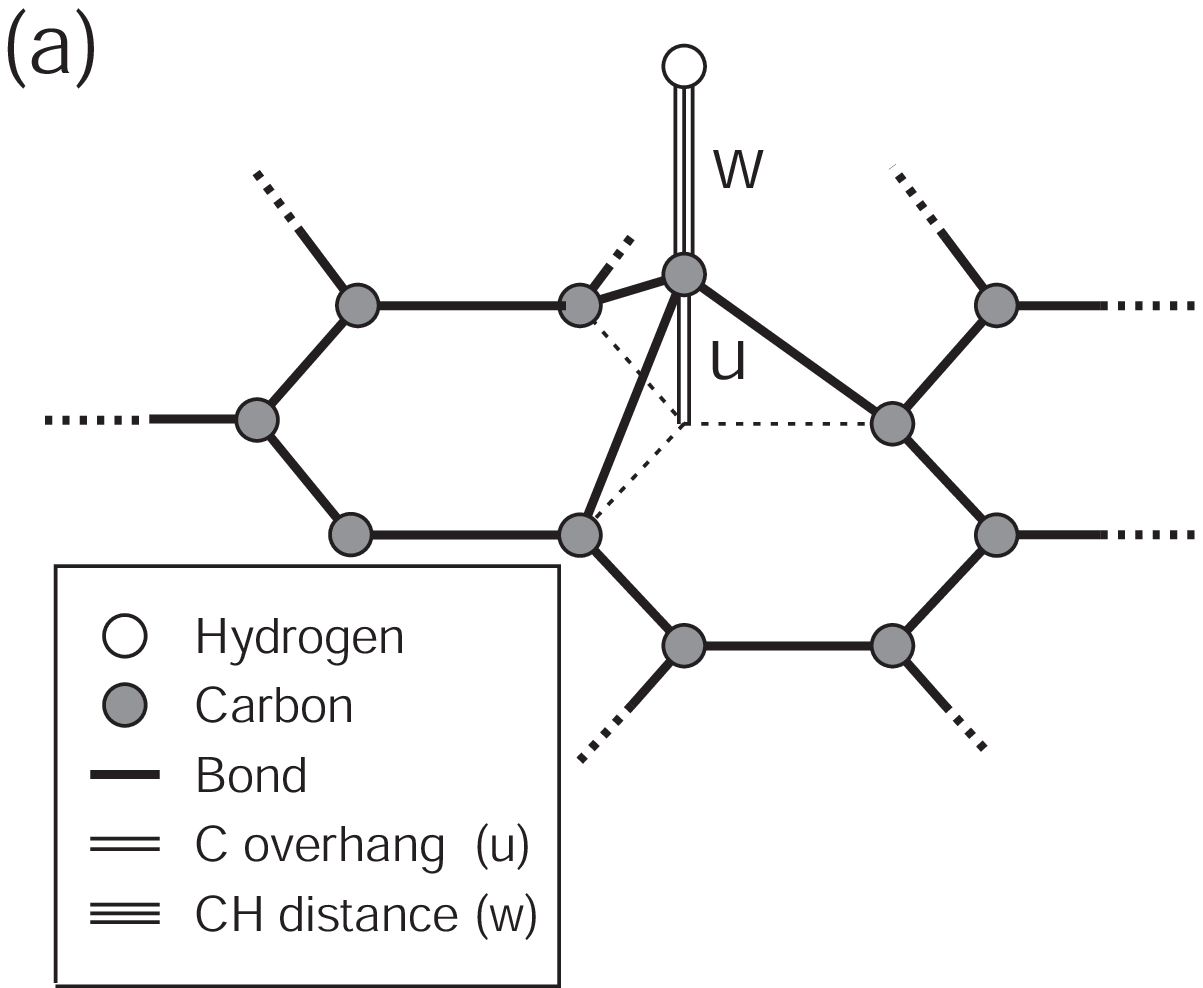}} &
		\resizebox{60mm}{!}{\includegraphics{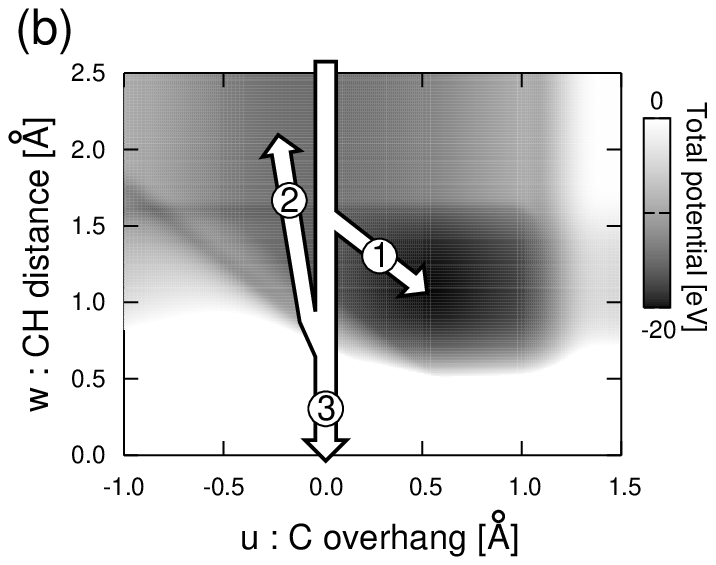}} \\
    \end{tabular}
\caption{%Three types of processes in the hydrogen-carbon absorption system.
(a) Schematic of a hydrogen atom fixed above a carbon atom showing the overhang from flat graphene ($x$)
 and the distance between the carbon atom and the incident hydrogen atom ($y$). 
(b) Schematic of reaction paths in the hydrogen--graphene collision process in total potential energy space. 
The white regions denote higher potential energy than $0$ eV.
Adsorption, reflection and penetration are denoted by arrows 1, 2 and 3, respectively.}
%The white region means higher potential area than black one.
%Arrows are paths of reaction.
%\caption{Three type process in the hydrogen-carbon absorption system.
%Meaning of x-y axis are explained in Fig(d).
%Arrows in Fig (a) to (c) are routes of the system.
%(a) Absorption process due to a low incident hydrogen energy in the potential map.
%(b) Reflection process when the incident energy is higher than the absorption process.
%(c) Penetration process due to higher incident energy than the reflection process.
%(d) Explanation of x-y axis of fig (a) to (c).
% The x axis means a distance from the plane of a flat graphene to the carbon atom
% which bond with the incident hydrogen atom.
% The y axis means a distance between the incident hydrogen atom and the the carbon atom.}
\label{fig:fig2}
\end{figure}

\section{Results and Discussion}
In the present simulations of incident hydrogen energies $E_{\rm I}$ of less than 100~eV,
 the hydrogen atoms were found to be absorbed (or reflected) by the graphenes
 without destroying the graphenes.
%It was expected that the hydrogen atoms would destroy the graphene.
As shown in Fig. \ref{fig:fig1}(a),
 the simulations suggest that the hydrogen adsorption rate on graphene
 is independent of the graphene temperature. 
This result is not consistent with experimental facts[\cite{Mech,Roth}]. 
However, the temperature defined in this simulation ($T_s$)
 is not the same as the temperature measured in experiments ($T_e$)
 due to the omission of inter-layer interaction in the present simulation.
The graphenes form a domain structure by the inter-layer interaction
 in the carbon wall.
%In our simulation, each graphene layer is placed independently without inter-layer interaction. 
%That is, graphene layers interact to form a domain structure in experiments,
% in many cases forming a `real' wall.
The experimental temperature $T_e$ therefore includes all the kinetic energy of the relative motion of
 each domain, each layer, and each atom, whereas $T_s$ accounts solely for the kinetic energy of atoms.

Figure \ref{fig:fig1}(b) shows that the hydrogen adsorption rate is dependent on $E_{\rm I}$.
%In order to understand this dependence,
The mechanism of adsorption can be classified into the following three types of processes:

{\bf Adsorption at $E_{\rm I} < 5$~eV: }
As the incident hydrogen atom approaches the graphene,
 the nearest carbon atom
% is pulled from the graphene plane,
% resulting in overhang from the flat graphene (Fig. \ref{fig:fig2}(a)). This 
overhangs spontaneously from the flat graphene
(Fig. \ref{fig:fig2}(a)). Because, this
 deformation of the graphene leads the total potential energy to the minimum point
 (arrow 1, Fig. \ref{fig:fig2}(b)). 
The kinetic energy converted from potential energy %absorption and $E_{\rm I}$ 
is diffused into the entire atoms in the graphene through the overhanging carbon atom. 
The incident atom thus loses kinetic energy and is trapped.

% and becomes like a hill shown by Fig. \ref{fig:fig2}(a).
%It is because the system moves to the stabilization point on potential energy 
% like an arrow 1 in Fig. \ref{fig:fig2}(b).
%Since a incident energy and energy gained by binding are diffused by the overhanging carbon atom
%  to other carbon atoms, 
% the incident atom is absorbed.

%Overhanging means that the carbon atom moves to the incident hydrogen atom
%and only that place become like a hill in a flat graphene.
%That reason is explained by potential map which is Figure \ref{fig:fig2}.
%The x-axis means the height that carbon atom overhangs,
%and the y-axis means the distance between the incident hydrogen atom and overhanging carbon atom.
%Those values are expressed as a doublet and a triplet in Fig. \ref{fig:fig2}(d).
%When the incident hydrogen atom does not interact with carbon atoms,
%the system is in which $x=0, y > 1.8$ in Fig. \ref{fig:fig2}(a).
%The incident hydrogen atom advance like the first arrow.
%When y value became less than 1.8 \AA, chemical reaction starts.
%The system want to advance to lower potential point in accordance with potential curve.
%For that reason, not only does the incident hydrogen atom approach the carbon atom,
%but also the carbon atom must overhang.
%In other word, the system advance like the second arrow.
%Because the obtained kinetic energy diffuse to other carbon atoms,
%the hydrogen atom can go away no longer. that is meaning absorption.

{\bf Reflection at $5$~eV~$< E_{\rm I} < 50$~eV: }
The velocity of the incident atom is
% too high under higher-energy conditions
% to pull the carbon atom into overhang,
  so high under high-energy conditions that spontaneous overhanging of the carbon atom can not occur,
 and the total potential does not pass through the minimum point. 
The incident atom bounces off this potential barrier (arrow 2, Fig. \ref{fig:fig2}(b)) and the 
%is reflected by the white region in Fig. \ref{fig:fig2}(b)).
incident atom is reflected. The adsorption rate under such conditions is therefore low.
%This process corresponds to  .
%Until starting chemical reaction, similar to the first type process,
%the system advance like first arrow Fig. \ref{fig:fig2}(b).
%However, the velocity of the incident hydrogen atom is too fast for the carbon atom to overhang
% while interaction with each other.
%Therefore, the system goes straight on like the second arrow.
%The incident hydrogen atom hits the potential wall which is higher potential area painted white,.
%The incident hydrogen atom is reflected by that wall soon.
%Thus, the incident hydrogen atom is hardly trapped
% since the carbon atom can not overhang.
%As a result, the adsorption rate decreases.

{\bf Penetration at $E_{\rm I} > 50$~eV: }
%In this case, the carbon atom can not overhang again.
%However, 
When the energy $E_{\rm I}$ is greater than the potential barrier,
%Therefore, the reaction path is drawn by .
the incident atom penetrates the graphene (arrow 3, Fig. \ref{fig:fig2}(b)).
%the path of incident atom is curved 
The incident atom gradually loses kinetic energy by repulsion of the carbon atom as it penetrating several graphene layers. The reaction thus eventually changes into a reflection process. After repeated bouncing between graphene layers, the incidence atom is trapped.
%the reaction becomes the absorption process.
%The hydrogen atom repeats bounding 
%Finally, 
%and  the incidence atom is trapped by some graphene.

\section{Conclusions}

%The chemical reaction of hydrocarbon 
The collision processes between incident hydrogen and graphene at energies of less than 100~eV
 were investigated by CMD simulations using a modified REBO potential. 
It was found that hydrogen does not destroy the graphene at these energies,
 but is instead absorbed or reflected by the graphene. 
The rate of hydrogen adsorption was shown to be dependent on the incident energy
 and not on the graphene temperature. 
The collision behavior was classified into the three processes;
 absorption of hydrogen in graphene ($E_{\rm I} < 5$~eV),
 reflection by graphene ($5$~eV $< E_{\rm I} < 50$~eV),
 and penetration through one or more graphenes ($E_{\rm I} > 50$~eV).


\begin{thereferences}{19}
	\bibitem{Nakano}
		Nakano, T., Kudo, H., Higashijima, S., Asakura, N., Takenaga, H., Sugie, T. and Itami, K. 
%     {Measurement of the chemical sputtering yields of ${\rm CH_4/CD_4}$
%		 and ${\rm C_2H_x/C_2D_x}$ at the carbon divertor plates of JT-60U.}
		2002 {\it Nucl. Fusion} {\bf 42}, 689. %--696.

	\bibitem{rebo}
		Brenner, D. W., Shenderova, O. A., Harrison, J. A., Stuart, S. J., Ni, B. and Sinnott, S. B. 
%		{A second-generation reactive empirical bond order (REBO) potential energy
%		expression  for hydrocarbons.}
		2002 {\it J. Phys. Condens. Matter} {\bf 14}, 783. %--802.
	\bibitem{Morse}
		Morse, P. M. 
%		{Diatomic molecules according to the wave mechanics. II. vibrational levels.}
		1929 {\it Phys. Rev.} {\bf 34}, 57. %--64.
	\bibitem{Abell}
		Abell, G. C. 
%		{Empirical chemical pseudopotential theory of molecular and metallic bonding.}
		1985 {\it Phys. Rev. }B {\bf 31}, 6184. %--6196.
	\bibitem{Tersoff}
		Tersoff, J. 
%		{New empirical approach for the structure and energy of covalent systems.}
		1987 {\it Phys. Rev. }B {\bf 37}, 6991. %--6999.
%	\bibitem{Garcia}
%		Garcia-Rosales, C. and Roth. J. 1992
%%		{New empirical approach for the structure and energy of covalent systems.}
%		{\it J. Nucl. Mater. } {\bf 196--198}, 573--.
	\bibitem{Mech}
		Mech, B. V., Haasz, A. A. and Davis, J. W. 
%		{Chemical erosion of pyrolytic graphite by low-energy H^{+} and D^{+} impact.}
		1997 {\it J. Nucl. Mater. } {\bf 241--243}, 1147. %--1151.
	\bibitem{Roth}
		Roth, J., and Garcia-Rosales, C. 
%		{Analytic description of the chemical erosion of graphite by hydrogen ions.}
		1996 {\it Nucl. Fusion } {\bf 36}, 1647. %--1659.
\end{thereferences}
\end{document}